\PassOptionsToPackage{finalnew}{trackchanges}
\documentclass[draft]{agujournal2018}
\usepackage{apacite}
\usepackage{url} 
\usepackage{lineno}
\usepackage[utf8]{inputenc}
\draftfalse
\journalname{Space Weather}

\usepackage{multirow}

\usepackage[separate-uncertainty=true, quotient-mode=fraction]{siunitx}
\DeclareSIUnit{\AU}{AU}
\DeclareSIUnit{\solarradius}{R_{{\mbox{$\odot$}}}}
\DeclareSIUnit{\sol}{sol}

\soulregister\ref7
\soulregister\cite7
\soulregister\citet7
\soulregister\citep7

\begin{document}
	
\title{Tracking and validating ICMEs propagating towards Mars using STEREO Heliospheric Imagers combined with 
	Forbush decreases detected by MSL/RAD}
\authors{
	Johan L. Freiherr von Forstner\affil{1},
	Jingnan Guo\affil{2,1},
	Robert F. Wimmer-Schweingruber\affil{1},
	Manuela Temmer\affil{3}, 
	Mateja Dumbović\affil{3},
	Astrid Veronig\affil{3},
	Christian Möstl\affil{4},
	Donald M. Hassler\affil{5},
	Cary J. Zeitlin\affil{6},
	Bent Ehresmann\affil{5}}
\affiliation{1}{Institute of Experimental and Applied Physics, University of Kiel, Kiel, Germany}
\affiliation{2}{School of Earth and Space Sciences, University of Science and Technology of China, Hefei, China}
\affiliation{3}{Institute of Physics, University of Graz, Graz, Austria}
\affiliation{4}{Space Research Institute, Austrian Academy of Sciences, Graz, Austria}
\affiliation{5}{Southwest Research Institute, Boulder, CO, USA}
\affiliation{6}{Leidos, Houston, TX, USA}

\correspondingauthor{Jingnan Guo}{guo@physik.uni-kiel.de}

\begin{keypoints}
	\item 149 ICMEs propagating towards Mars are studied, combining data from STEREO Heliospheric Imagers with Forbush 
	decreases at MSL/RAD
	\item 45 ICMEs can be clearly associated with a Forbush decrease at MSL/RAD. Many others are uncertain due to e.g. 
	CME-CME interaction.
	\item Arrival times predicted from HI data agree with RAD data with a standard deviation of $\sim$ 17 to 19 hours
\end{keypoints}

\section*{Abstract}
The Radiation Assessment Detector 
(RAD) instrument onboard the Mars Science Laboratory (MSL) mission's Curiosity rover has been measuring galactic cosmic 
rays (GCR) as well as solar energetic particles (SEP) on the surface of Mars for 
more than 6 years since its landing in August 2012. The observations include a large number of Forbush decreases (FD) 
caused by interplanetary coronal mass ejections (ICMEs) and/or their associated shocks shielding away part of the GCR 
particles with their turbulent and enhanced magnetic fields while passing Mars.

This study combines MSL/RAD FD measurements and remote tracking of ICMEs using the STEREO Heliospheric Imager (HI) 
telescopes in a statistical study for the first time. The large dataset collected by HI makes it possible to analyze
149 ICMEs propagating towards MSL both during its 8-month cruise phase and after its landing on Mars. We link 45 of the 
events observed at STEREO-HI to their corresponding FDs at MSL/RAD and study the accuracy of the ICME arrival time at 
Mars predicted from HI data using different methods.

The mean differences between the predicted arrival times and those observed using FDs range from 
\SIrange{-11}{5}{\hour} for the different methods, with standard deviations between 17 and 20 hours. These values for 
predictions at Mars are very similar compared to other locations closer to the Sun, and also comparable to the 
precision of some other modeling approaches.

\section{Introduction}\label{chp:introduction}
Coronal Mass Ejections (CMEs), clouds of magnetized plasma expelled from the Sun, often at high speeds, are one of 
the main concerns of space weather research. The passage of CMEs at Earth can cause geomagnetic storms 
\citep[e.g.][]{Cane-2000-ICME-geomagnetic}, which are 
severe disruptions of the terrestrial magnetic field that can in some cases have serious impact on infrastructure on 
the surface of Earth, such as damaging electricity grids 
\citep{Oughton2017-CMEimpact,Boteler-1998-Geomagnetic-effects}.
Additionally, shocks driven by fast CMEs are believed to be one of the phenomena responsible for the acceleration of 
solar energetic particles (SEPs, \add{see e.g. \protect\citet{Reames-2013}}), which may cause radiation damage to 
spacecraft, aircraft and astronauts. 
Consequently, the observation, modeling and, eventually, forecasting of 
CMEs and their interplanetary counterparts (ICMEs) as well as their impacts on Earth have been important topics in the 
space weather community in the last decades.

CMEs are usually detected remotely using coronagraph instruments, while their interplanetary counterparts can be 
observed \textit{in situ} by their signatures in the interplanetary magnetic field, as well as plasma parameters such 
as the solar wind speed, density and temperature \citep[cf.][]{Zurbuchen-2006-insitu-signatures}. Additionally, ICMEs 
impact the galactic cosmic ray (GCR) flux in the form of short-term decreases, first observed by \citet{Forbush-1937} 
and
\citet{Hess-1937}, and later named Forbush decreases (FDs). Since then, numerous authors have studied these effects, 
which are caused by the magnetic field structure of the ICME and/or its preceding shock shielding parts of the incoming 
GCR away from the measurement location. The decrease usually takes less than a day while the recovery period of the GCR 
flux back to its previous level can be much longer (about one week). In the case where both a shock/sheath and the ICME 
ejecta pass the measurement location, the FD can show a two-step structure, as described by e.g. \citet{Cane-2000}.
FD measurements are suitable for the detection of the arrival time of ICMEs as their onset time usually matches very 
closely to the corresponding solar wind structure \citep{Cane-1996,Dumbovic-2011}, and multiple researchers have 
previously used FDs for this purpose in cases where solar wind and magnetic field measurements are not available 
\citep[e.g.][]{Moestl-2015,Lefevre-2016,Vennerstrom-2016}, as was the case at Mars until the MAVEN mission arrived in 
September 2014 --- and even MAVEN does not continuously measure the upstream solar wind due to its elliptic orbit that 
regularly enters Mars's magnetosphere. \add{\citet{Witasse-2017} have also used Forbush decreases to observe 
the same ICME at multiple locations in the heliosphere out to {\SI{9.9}{\AU}}.}

Similar decreases in the GCR flux can also be caused by 
stream interaction regions (SIRs), which are the regions where fast and slow solar wind interact. SIRs often repeat 
for several solar rotations when the coronal hole structures at the solar surface producing these high speed streams 
are long-lived \citep[see e.g.][]{Heinemann-2018}. These recurrent GCR decreases are also called FDs by some authors, 
but in this work, we focus on ICME-caused FDs, as we are using them to investigate the propagation of ICMEs.

\begin{figure}
	\centering
	\includegraphics{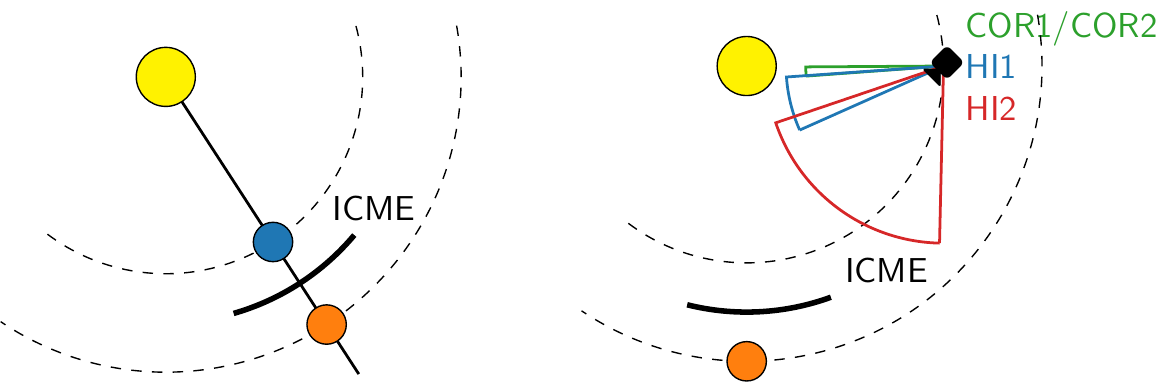}
	\caption[Opposition phases and STEREO SECCHI observations]{Cartoon comparison of the opposition phase constellation 
	to the observation of ICMEs with the STEREO SECCHI instruments. Observations of the same ICME at Earth and Mars are 
	only possible within a small longitudinal separation between the two planets \citep[left, see][]{Forstner-2018}, 
	while the STEREO-HI telescopes allow a continuous remote tracking of ICMEs in a wider range of directions 
	(right, this study)}
	\label{fig:introduction_oppositions_cartoon}
\end{figure}

Considering the increased interest in the exploration of Mars in the recent times, with multiple ongoing robotic 
missions and plans for human missions in the future, it becomes important to study the effects of radiation and space 
weather on Mars as well. In our previous work 
\citep{Forstner-2018}, we presented the first statistical studies of ICMEs arriving at Mars, using 
data from the \textit{Mars Science Laboratory} (MSL) mission's \textit{Radiation Assessment Detector} (RAD) instrument, 
which can detect FDs caused by ICMEs passing Mars. The study was based on \textit{in situ} observations of 
the same ICMEs, first at Earth or one of the STEREO spacecraft (i.e., at a radial distance of about \SI{1}{\AU} from 
the Sun), and then at Mars (which has a radial distance of about \SI{1.5}{\AU}) yielding the result that most ICMEs in 
our sample continued to decelerate slightly beyond \SI{1}{\AU}, dragged by the slower surrounding solar wind. The 
amount of deceleration and the ICME speed relative to the ambient solar wind were also found to have a tendency to 
correlate, but the statistical significance was limited due to the small number of 15 events that could be studied 
during close alignment of Mars and Earth or STEREO (cf. Figure \ref{fig:introduction_oppositions_cartoon}, left panel).

To enable a more complete observation-based study of ICMEs propagating towards Mars, we turn to remote observations in 
the current study, in particular those made possible by the Heliospheric Imager (HI) telescopes on the STEREO 
spacecraft. They facilitate tracking of ICMEs all the way from 
the Sun to approximately \SI{1}{\AU} in a wide range of directions, thus making the study less dependent on a certain 
constellation of planets and spacecraft (Figure \ref{fig:introduction_oppositions_cartoon}, right panel).

In this study, we will combine data from the STEREO-HI instruments with MSL/RAD observations to investigate ICMEs and 
FDs at Mars in more detail as well as to validate the accuracy of determining the ICME arrival at MSL 
using STEREO-HI data.

\section{Data and methods}\label{chp:data_methods}
\subsection{The MSL/RAD instrument}
\label{sec:mslrad}

Since the \textit{Curiosity} rover of NASA's \textit{Mars Science Laboratory} (MSL) mission 
\citep{Grotzinger-2012-MSL} landed on Mars on August 6, 2012, its \textit{Radiation Assessment Detector} 
\citep[RAD,][]{Hassler-2012-MSLRAD} instrument, built in a cooperation between Kiel University, German Aerospace 
Center (DLR) and Southwest Research Institute (SwRI), has been continuously measuring the particle radiation 
environment on the surface of Mars. RAD can detect neutral and charged particles using a setup of 6 detectors, named A 
through F. For two of the detectors (B and E), the dose rate contributed from all particles observed in the detector is 
also measured.

The radiation measured on the surface of Mars consists of primary galactic cosmic rays (GCR) and solar energetic 
particles (SEP) as well as secondary particles created when the primary radiation interacts with the Martian atmosphere 
\citep[e.g.][]{Guo-2018-Modeling}. 
The daily variation of atmospheric pressure causes a diurnal pattern in the dose rate measured at MSL/RAD 
\citep{Rafkin-2014}. Similar to neutron monitors on Earth and other cosmic ray detectors in deep space, RAD can be used 
for detecting FDs in the GCR. Due to the larger geometric factor, the dose rate in the E detector, a plastic 
scintillator, is best used for this purpose. To simplify the detection of FDs in the RAD data, the dose rate 
measurements are processed using a spectral notch filter described by \citet{Guo-2017-maven} to compensate for the 
diurnal variations.

RAD was also active during most of the time of the MSL rover's flight from Earth to Mars from December 2011 to 
July 2012 (the so-called \textit{cruise phase}). Without the Martian atmosphere around it, part of the RAD view cone 
was only very lightly shielded during this period \citep{Zeitlin-2013-cruise} and thus observed a 
different range of energies in the primary GCR spectrum and a higher number of SEP events. Besides, FDs were also 
detected during the cruise phase \citep{guo2015cruise}.

\subsection{The STEREO Heliospheric Imagers}
\label{sec:stereo_hi}

The Solar TErrestrial RElations Observatory (STEREO) mission \citep{Russell-2008-STEREO} was launched in 2006. Its 
two spacecraft, STEREO A (\textit{Ahead}) and STEREO B (\textit{Behind}) enabled a stereoscopic view of the Sun and 
inner heliosphere for the first time, with one of the main objectives being the study of CMEs and their impact on Earth.
The Sun Earth Connection Coronal and Heliospheric Investigation instrument suite 
\citep[SECCHI,][]{Howard-2008-SECCHI} on STEREO consists of multiple telescopes 
observing the Sun and heliosphere with different fields of view and wavelengths. The Heliospheric Imagers HI1 and HI2 
are white-light telescopes and have the largest field of view --- combined ranging from \SIrange{4}{88.7}{\degree} on 
one side of the Sun. This provides an excellent opportunity to observe ICMEs traveling from the Sun outward to 
\SI{1}{\AU} and beyond.

Connection to the STEREO B spacecraft was lost on October 1, 2014, a few months before its solar conjunction. So after 
this date, data is only available from STEREO A.

\subsection{Reconstruction of ICME kinematics from STEREO-HI data}
\label{sec:hi_kinematics_reconstruction}

As ICME propagation is a three-dimensional phenomenon, it is not trivial to derive the trajectory 
from just one (or, for ICMEs seen by both STEREO spacecraft, two) series of 2D images from STEREO-HI. The analysis of 
HI images is therefore typically based on the identification of ICMEs in J-Map (time-elongation map) diagrams\change{ 
\protect\citep[as described by][]{Sheeley-1999-JMap,Rouillard-2008-JMap}}{, developed by \citet{Sheeley-1999-JMap} for 
the analysis of coronagraph images and applied to STEREO-HI by \citet{Rouillard-2008-JMap} and \citet{Davies-2009},} 
combined with a number of different single-spacecraft fitting or multispacecraft triangulation methods to reconstruct 
the ICME kinematics.

The time-elongation profile identified in J-Maps is converted to the ICME's radial distance $r(t)$ from the Sun based 
on assumptions about the shape of the ICME and its appearance in the HI images. Different approaches for this 
conversion can be separated into two types: Single-spacecraft reconstruction methods are based on images from just one 
of the STEREO spacecraft, while multispacecraft approaches use data from both 
spacecraft to further constrain the parameters of the ICME trajectory 
\citep[e.g.][]{Liu-2010a-triangulation, Liu-2010b-triangulation,Lugaz-2010-TAS}. 

\begin{figure}
	\centering
	\includegraphics{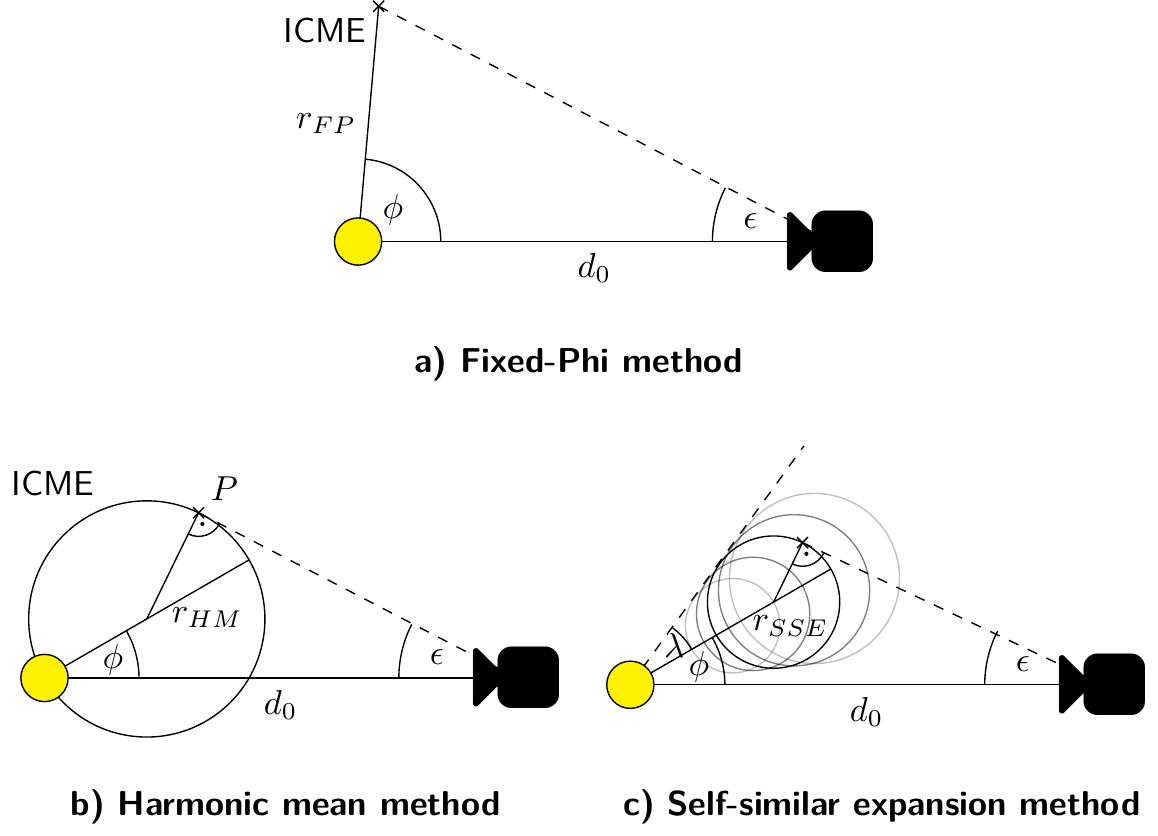}
	\caption{Single-spacecraft reconstruction methods for Heliospheric Imaging observations: a) Fixed $\phi$ 
	\citep{Kahler-2007-FixedPhi}, b) Harmonic mean \citep{Lugaz-2009-HarmonicMean}, and c) Self-similar expansion 
	\citep{Davies-2012-SSE,Lugaz-2010-SSE}.}
	\label{fig:geometric_methods}
\end{figure}

\add{These multispacecraft triangulation methods have been used to study many ICMEs propagating towards Earth, and they 
make it 
possible to relax some of the assumptions that need to be made when using only single-spacecraft methods. This allows 
for a detailed study of the ICME's speed profile within the HI field of view 
\protect\citep[e.g.][]{Liu-2013,Liu-2016}}. 
\change{As}{However,} the orbits of the STEREO spacecraft are optimized for observing Earth-directed structures and 
thus multi-spacecraft HI observations are often not possible for ICMEs propagating towards Mars\change{,}{. Also, our 
study includes some events after 2014, where only STEREO A is available. Therefore,} we will focus on the most common 
single-spacecraft methods here, which are shown in Figure \ref{fig:geometric_methods}.

One of the \change{earliest}{simplest} single-spacecraft reconstruction methods \change{was}{is} the Point-P method 
\citep[PP,][]{Howard-2006-PointP}, where the ICME is regarded as an expanding circular front centered around the Sun. 
On the contrary, the Fixed-$\phi$ model 
\change{\protect\citep[FP,][]{Kahler-2007-FixedPhi}}{\protect\citep[FP,][]{Sheeley-1999-JMap,Kahler-2007-FixedPhi}} 
reduces 
the shape of the ICME to a single point moving away from the Sun radially with a fixed longitudinal separation $\phi$ 
from the observer. The Harmonic Mean method \citep[HM,][]{Lugaz-2009-HarmonicMean} is a middle ground between these two 
extremes, where the circular ICME has one edge fixed at the Sun instead of being centered around it, which is 
equivalent to calculating the harmonic mean of the PP and FP results. \add{Appendix B of \cite{Liu-2010b-triangulation} 
contains a more detailed description of these three methods.}

Finally, the \add{more recent} Self-similar-expansion method
\citep[SSE,][]{Davies-2012-SSE,Lugaz-2010-SSE} introduces a second parameter $\lambda$ for the half-width of the 
ICME, which can be seamlessly adjusted between the two edge cases for $\lambda = \SI{90}{\degree}$ ($\Rightarrow$ HM) 
and $\lambda = \SI{0}{\degree}$ ($\Rightarrow$ FP).

The FP, HM and SSE methods are usually combined with fitting algorithms (FPF, HMF and SSEF) to determine parameters 
such as the longitudinal propagation direction of the CME under the assumption of a constant speed. As the leading edge 
of the visible structure is marked in the J-Map, $r(t)$ and the resulting speed $v$ are believed to most likely 
correspond to the shock front, if present, and otherwise the front of the ICME ejecta.

\subsection{The HELCATS catalogs}
\label{sec:helcats_catalogs}
On its website at \url{https://www.helcats-fp7.eu}, the \textit{Heliospheric Cataloguing, Analysis and Techniques 
Service} EU project \citep[HELCATS,][]{helcats_higeocat} has 
collected a large number of ICMEs observed with the STEREO-HI instruments. Their \textit{HIGeoCat} CME kinematics 
catalog (version 5) contains 1459 ICMEs, each one supplemented with the associated J-Map and 
a time-elongation profile extracted from it by manual selection. Results for ICME speeds and propagation 
directions derived using the FPF, HMF and SSEF methods are also provided (under the assumption of a constant speed $v$, 
longitude $\phi$, and in the SSE case, a fixed half-width of $\lambda = \SI{30}{\degree}$).

The \textit{ARRCAT} arrival catalog contains a list of predicted \textit{in situ} arrival times of ICMEs at different 
planets and spacecraft including MSL, based on the events in the \textit{HIGeoCat} and currently (version 01) updated 
until the end of September 2014, where STEREO B data ends. As described by \citet{Moestl-2017-HelcatsHSO}, the 
arrival time predictions were calculated based on the SSEF30 method results from HIGeoCat, extrapolating the trajectory 
up to the respective location based on the constant speed $v$. The calculation also includes the correction described 
by \citet{Moestl2013-arrival-correction} to account for the SSE geometry at locations that are not directly hit by 
the ICME apex.

\citet{Moestl-2017-HelcatsHSO} also compared the ARRCAT data for multiple locations in the inner heliosphere with in 
situ plasma and magnetic field data to check the accuracy of the predicted arrival times. However, their study does not 
include MSL or other spacecraft at Mars.

\section{Results and Discussion}\label{chp:results_discussion}
\subsection{ICMEs observed by STEREO-HI and their arrival times at MSL/RAD}
\label{sec:arrivaltimes}

To analyse ICMEs that arrived at MSL, we first select candidate events from the HELCATS HIGeoCat by requiring the 
propagation direction (heliospheric longitude, determined using the SSEF30 method) to be within $\pm\SI{30}{\degree}$ 
of MSL's longitude at that time. Although the longitude \change{may depend}{depends} considerably on the geometry used 
for fitting the HI data \add{\citep{Lugaz-2009-SoPh,Liu-2013,Liu-2016}}, we use the modeled direction from SSE as it is 
the most advanced of the three methods available in the HIGeoCat. Also, this selection 
criterion is equivalent to the one used for the HELCATS ARRCAT catalog and the corresponding paper by 
\citet{Moestl-2017-HelcatsHSO}, making it possible to compare to their results afterwards (see 
Section \ref{sec:arrival_comparison}). The selection yields 149 ICME events between the beginning of MSL's cruise phase 
in November 2011 and the current end of the HIGeoCat catalog (version 5) in November 2017. 31 of these events were 
observed with both STEREO spacecraft according to the HELCATS HIJoinCat catalog, the remaining 118 ICMEs were only 
detected by one of them. \add{For events where observations from both spacecraft were available, we use the spacecraft 
data which predict an arrival at MSL based on the SSEF30 fitting. If both STEREO observations predict the ICME to 
arrive at MSL, we use STEREO A data as a preference.}

\begin{figure}
	\centering
	\includegraphics{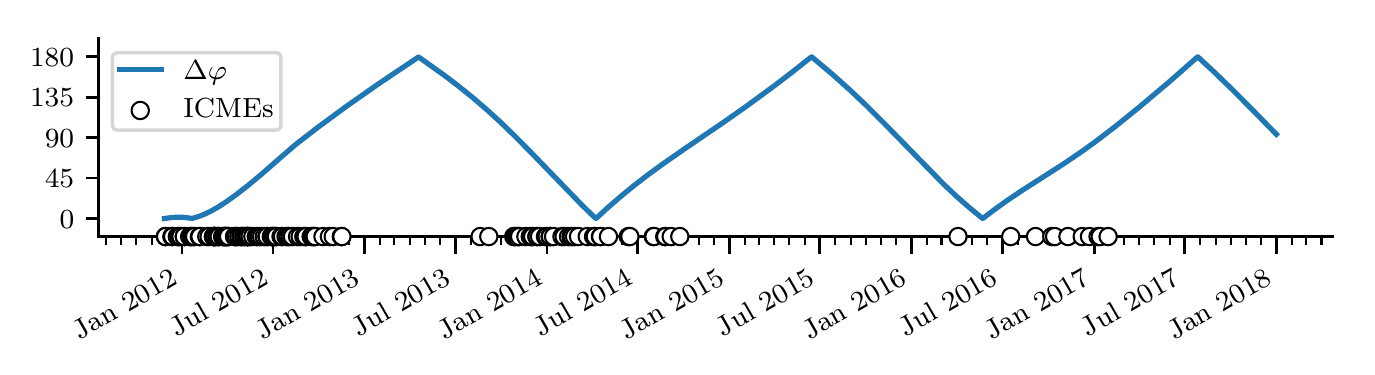}
	\caption[Distribution of events in the catalog]{Time distribution of the 149 ICMEs propagating towards MSL 
		$\pm\SI{30}{\degree}$ from the HELCATS catalog, plotted together with the longitudinal separation $\Delta 
		\varphi$ 
		between Earth and MSL. During periods with large $\Delta\varphi$, STEREO-HI cannot see ICMEs propagating 
		towards 
		MSL, as it is only looking at one side of the Sun.}
	\label{fig:helcats_events_distribution}
\end{figure}

A plot showing the time distribution of the 149 events is displayed in Figure \ref{fig:helcats_events_distribution}. It 
can easily be seen that the ICMEs are not evenly distributed, but rather concentrated into three discrete periods. The 
reason for this is that in contrast to the coronagraph instruments, the heliospheric imagers' fields of view are 
limited to one side of the Sun, and the STEREO spacecraft are always pointed in a way that the side of the Sun at which 
HI looking is the one where Earth is located. So when e.g. Earth is on the right side of the Sun as seen from one of the
STEREO spacecraft while Mars is on the left side, ICMEs propagating towards Mars cannot be seen using the HI 
instruments on this spacecraft. The last period in 
2016--17 has fewer events, which is related to the solar cycle approaching its minimum as well as the loss of 
STEREO B data since 2014.

\begin{figure}
	\centering
	\scalebox{0.85}{\includegraphics{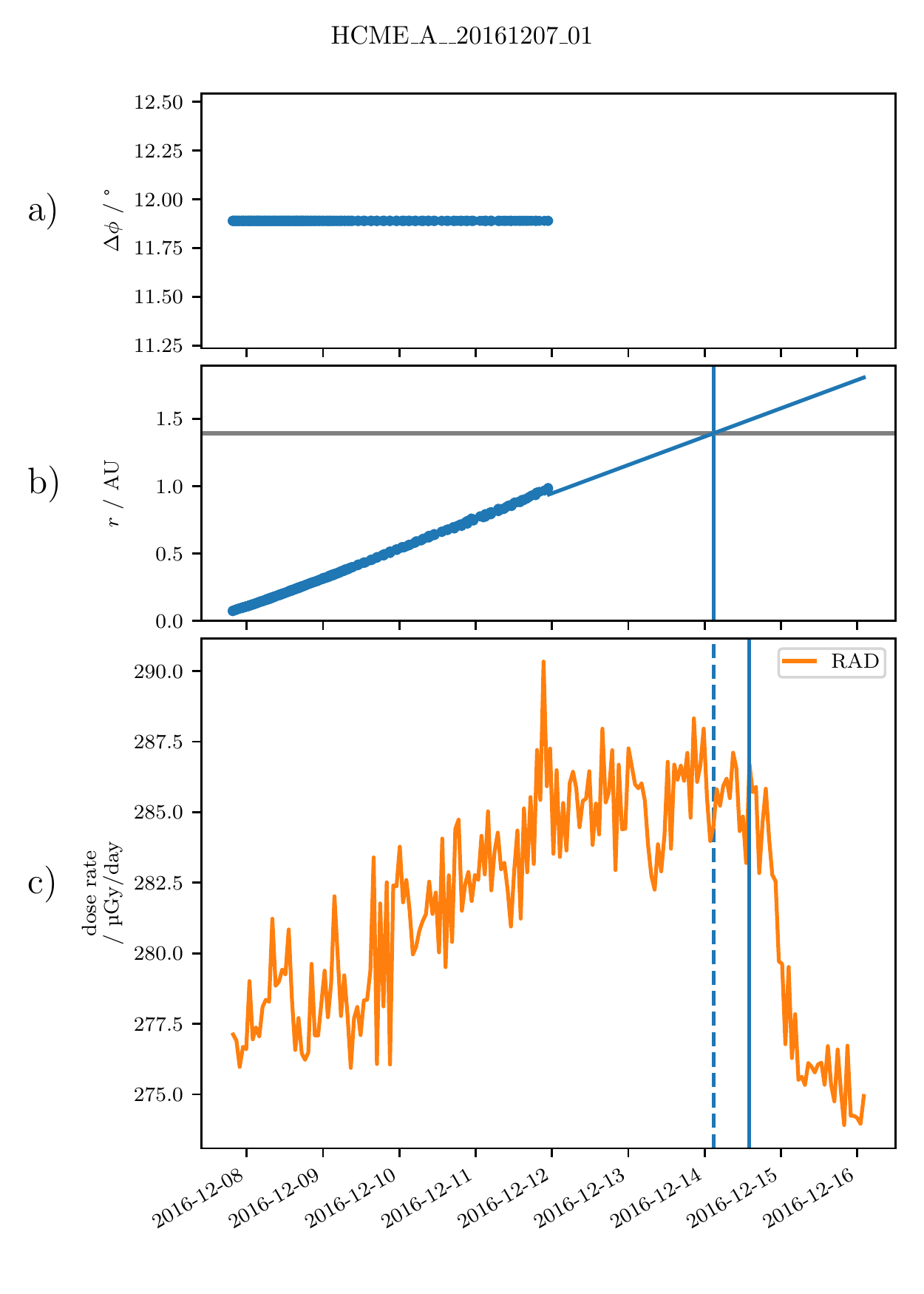}}
	\caption[Example comparison of STEREO-HI arrival prediction and FD]{An example of the type of plot used 
		to compare the ICME arrival times at Mars that were predicted using STEREO-HI observations and the 
		onset time of the corresponding FD detected at MSL/RAD. This plot is for the
		``HCME\_A\_\_20161207\_01'' event from the 
		HELCATS catalog. \change{The lower panel}{Panel c)} shows the MSL/RAD dose rate measurements where the FD onset 
		is marked with a 
		solid blue line. \change{Above that}{In panel b)}, the trajectory of the ICME in the direction of MSL (radial 
		distance from the Sun 
		over time) is plotted, which was calculated from the time-elongation data using the SSEF30 method and then 
		extrapolated assuming the launch time and the constant speed that result from the SSEF30 fitting procedure.
		\add{The blue vertical line in this panel marks the time where the extrapolation intersects Mars's radial 
			distance from the Sun, i.e. the predicted arrival time at Mars, and this line was extended into panel c) 
			as a dashed line.} \change{The upper two 
			panels show}{Panel a) shows} the longitudinal separation between Mars and the ICME apex (in this case 
			constant, because it is determined from the SSEF30 fit).}
	\label{fig:elongation_plot_example}
\end{figure}

For each of the 149 events, a plot similar to Figure \ref{fig:elongation_plot_example} was constructed, showing time 
series of the ICME's radial distance from the Sun, which were derived using the SSE geometry and the longitude derived 
from the SSEF30 method results (\change{bottom panel}{panel b)}), the longitudinal 
separation between the CME apex and MSL's location (\change{top panel}{panel a)}) as well as the MSL/RAD dose rate 
processed using 
the notch filter described in Section \ref{sec:mslrad} (\change{middle panel}{panel c)}). The time-distance plot was 
corrected based on 
the longitudinal separation between the ICME apex and MSL using the 
equation from \citet{Moestl2013-arrival-correction} so that it shows the radial distance of the part of the ICME in the 
SSE geometry that hits MSL, therefore resulting in a slightly lower speed. The predicted ICME arrival time at MSL could 
then be calculated based on the $r(t)$ trajectory and was marked with a vertical line in both the $r(t)$ and RAD dose 
rate panels.

It needs to be stated that these ``predictions'' were done on the basis of post-event analysis using science data and 
not real-time beacon data. 
However, as ICMEs are usually only visible in HI images up to about \SI{1}{\AU} or less, the predictions at Mars are 
probably comparable to what could have been done with real time data in most cases --- in contrast to the ARRCAT 
predictions at locations closer to the Sun \citep{Moestl-2017-HelcatsHSO}, which are partly based on observations that 
would only have been available after the ICME arrival. A study using real time STEREO-HI data was conducted by 
\citet{Tucker-Hood-2015-HI-Realtime}.

Based on the predicted arrival time at MSL/RAD, a corresponding FD in the MSL/RAD data was searched for. In general, we 
used a time window of at most $\pm\SI{2.5}{day}$ around the predicted arrival time to make sure that the FD onset time 
was not too far off and thus maybe related to a completely different event. We also took care that series of events 
predicted to arrive in a certain order were matched to the FDs in the correct sequence if no interaction between the 
ICMEs is predicted, and that for CMEs also seen at Earth close to oppositions of MSL and Earth, the arrival at Mars is 
not earlier than the one at Earth. Although the time window might seem quite small for slow CMEs, a larger window would 
have caused more ambiguous cases with multiple candidate FDs.

The FD 
onset was marked to be the point in time where the GCR intensity reaches its maximum at the beginning 
of the FD. This makes the onset time a more well-defined quantity compared to just marking it ``by eye'', 
and was implemented by searching for the maximum within a $\pm\SI{4}{\hour}$ window around the onset time that was 
first manually selected.

In the process of marking the FD onset times, the 149 ICMEs were sorted into five categories, by looking 
at the RAD data as well as the ICME trajectories calculated from HI data and their extrapolations:

\begin{sidewaystable}
	\caption[Table of all 45 events where an arrival time at Mars could be determined]{Table of all 45 events where an arrival time at Mars could be determined (categories 1 and 2 from Section \ref{sec:arrivaltimes}). The first two columns show the IDs of the ICMEs in the HELCATS catalog for STEREO A and B observations, which also correspond to the date where the ICME was first observed in HI images\add{, and the third column shows the STEREO spacecraft that was used for applying the fitting methods}. The \change{third and fourth}{fourth and sixth} column show thelongitude $\phi$ and the speed $v$ of the ICME as determined by the HELCATS project using the SSEF30 method. \change{The following}{Additional} columns show the difference\add{s} between $\phi$ and the heliospheric longitude\add{s} of \add{the respective STEREO spacecraft (column 5) or} MSL \add{(column 7)} at that time as well as the \textit{in situ} arrival time (Forbush decrease onset) at MSL \add{and the transit time $T_\text{MSL}$ between the launch at the Sun and arrival at MSL}. The arrival time and longitudinal separation are also given for the arrival at Earth, if applicable.}
\scalebox{0.5}{
\begin{tabular}{cccSSSScSSc}
\hline
   HCME\_A\_\_\dots &   HCME\_B\_\_\dots &  SC  &   {$\phi_\text{SSE}$ / \si{\degree}} &   {$\Delta \phi_\text{STEREO}$ / \si{\degree}} &   {$v_\text{SSE}$ / \si{\kilo\meter\per\second}} &   {$\Delta \phi_\text{MSL}$ / \si{\degree}} &  $t_\text{MSL}$  &   $T_\text{MSL} / \si{\day}$ &   {$\Delta \phi_\text{Earth}$ / \si{\degree}} &  $t_\text{Earth}$  \\
\hline
        20111211\_01 &        20111211\_01 &  A   &                                    3 &                                            -27 &                                              452 &                                       -28.7 &  2011-12-17 01   &                          5.8 &                                           -27 &                    \\
        20111222\_01 &                    &  A   &                                  221 &                                            -24 &                                              320 &                                       -25.3 &  2011-12-28 08   &                          6.1 &                                           -24 &                    \\
                    &        20111226\_01 &  B   &                                  206 &                                             15 &                                              686 &                                        13.7 &  2011-12-30 01   &                          3.7 &                                            15 &                    \\
        20120123\_01 &        20120123\_01 &  A   &                                  207 &                                             20 &                                             1052 &                                        20.4 &  2012-01-24 21   &                          1.9 &                                            20 &                    \\
                    &        20120224\_01 &  B   &                                  100 &                                            -25 &                                              885 &                                       -17.6 &  2012-02-27 08   &                          3.2 &                                           -25 &   2012-02-26 23    \\
        20120310\_01 &        20120310\_01 &  A   &                                  266 &                                            -14 &                                             1447 &                                        -1.9 &  2012-03-12 21   &                          2.3 &                                           -14 &   2012-03-12 02    \\
        20120322\_01 &        20120322\_01 &  A   &                                  127 &                                             -3 &                                              469 &                                        13.1 &  2012-03-28 11   &                          6.5 &                                            -3 &   2012-03-26 20    \\
        20120415\_01 &        20120415\_02 &  B   &                                  133 &                                            -39 &                                              316 &                                       -13.4 &  2012-04-21 02   &                          5.9 &                                           -39 &                    \\
                    &        20120512\_01 &  B   &                                  159 &                                            -19 &                                              869 &                                        17.9 &  2012-05-15 12   &                          3.6 &                                           -19 &   2012-05-15 00    \\
        20120530\_01 &                    &  A   &                                  261 &                                            -30 &                                              686 &                                        15.5 &  2012-06-03 04   &                          3.6 &                                           -30 &   2012-06-02 07    \\
                    &        20120702\_01 &  B   &                                  163 &                                            -55 &                                              697 &                                         6.2 &  2012-07-06 07   &                          4   &                                           -55 &                    \\
        20120702\_01 &        20120702\_02 &  B   &                                  173 &                                            -44 &                                              398 &                                        17.4 &  2012-07-09 10   &                          7   &                                           -44 &                    \\
                    &        20120915\_01 &  B   &                                  223 &                                            -85 &                                              441 &                                        11   &  2012-09-21 17   &                          6.5 &                                           -85 &                    \\
                    &        20120918\_01 &  B   &                                  193 &                                            -84 &                                              412 &                                        13.1 &  2012-09-23 14   &                          6.1 &                                           -84 &                    \\
                    &        20120920\_01 &  B   &                                  141 &                                            -99 &                                              528 &                                        -1.6 &  2012-09-25 09   &                          5.2 &                                           -99 &                    \\
                    &        20120922\_01 &  B   &                                  111 &                                           -111 &                                              562 &                                       -13   &  2012-09-27 01   &                          5.5 &                                          -111 &                    \\
                    &        20120923\_01 &  B   &                                  121 &                                            -78 &                                              620 &                                        21.5 &  2012-09-28 10   &                          5   &                                           -78 &                    \\
                    &        20121022\_01 &  B   &                                   92 &                                            -95 &                                              427 &                                        16.3 &  2012-10-27 08   &                          5.7 &                                           -95 &                    \\
                    &        20121029\_01 &  B   &                                  357 &                                            -89 &                                              342 &                                        24.6 &  2012-11-05 16   &                          7.5 &                                           -89 &                    \\
                    &        20121115\_01 &  B   &                                  127 &                                           -110 &                                              349 &                                        10.3 &  2012-11-20 19   &                          6.7 &                                          -110 &                    \\
                    &        20121116\_01 &  B   &                                  108 &                                           -105 &                                              534 &                                        15.5 &  2012-11-20 06   &                          4.4 &                                          -105 &                    \\
        20130820\_01 &        20130820\_01 &  A   &                                  263 &                                            105 &                                              701 &                                       -22.5 &  2013-08-24 22   &                          5.1 &                                           105 &                    \\
        20131119\_01 &                    &  A   &                                   87 &                                             55 &                                              577 &                                       -24.4 &  2013-11-25 01   &                          5.8 &                                            55 &                    \\
        20140101\_01 &                    &  A   &                                  249 &                                             58 &                                              311 &                                         2.4 &  2014-01-10 11   &                          9.7 &                                            58 &                    \\
        20140114\_01 &                    &  A   &                                   65 &                                             46 &                                              655 &                                        -1.3 &  2014-01-18 09   &                          4.5 &                                            46 &                    \\
        20140130\_01 &                    &  A   &                                  197 &                                             37 &                                              451 &                                        -0.6 &  2014-02-06 18   &                          7.2 &                                            37 &   2014-02-05 18    \\
        20140204\_01 &                    &  A   &                                  111 &                                              9 &                                              832 &                                       -26.1 &  2014-02-08 14   &                          4.7 &                                             9 &   2014-02-07 20    \\
        20140213\_01 &        20140213\_01 &  A   &                                   42 &                                             58 &                                              661 &                                        28   &  2014-02-17 06   &                          4.4 &                                            58 &   2014-02-15 13    \\
        20140220\_01 &                    &  A   &                                  298 &                                             50 &                                              763 &                                        24.1 &  2014-02-24 04   &                          4   &                                            50 &   2014-02-23 14    \\
        20140225\_01 &                    &  A   &                                  198 &                                             13 &                                              661 &                                       -10.2 &  2014-03-01 10   &                          4.5 &                                            13 &   2014-02-28 02    \\
        20140322\_01 &                    &  A   &                                  244 &                                             34 &                                              798 &                                        24.9 &  2014-03-26 22   &                          4.6 &                                            34 &                    \\
        20140404\_01 &                    &  A   &                                   54 &                                              9 &                                              895 &                                         6.6 &  2014-04-06 16   &                          2.8 &                                             9 &   2014-04-05 11    \\
                    &        20140407\_01 &  B   &                                  335 &                                            -26 &                                              535 &                                       -26.5 &  2014-04-14 04   &                          7   &                                           -26 &   2014-04-11 16    \\
        20140418\_01 &                    &  A   &                                  218 &                                             -3 &                                             1265 &                                         1.8 &  2014-04-20 16   &                          2.8 &                                            -3 &   2014-04-20 09    \\
        20140418\_02 &                    &  A   &                                  235 &                                             22 &                                             1092 &                                        27.1 &  2014-04-22 02   &                          3.7 &                                            22 &   2014-04-20 17    \\
        20140504\_01 &                    &  A   &                                   15 &                                              4 &                                              866 &                                        16.9 &  2014-05-07 02   &                          3.4 &                                             4 &                    \\
                    &        20140801\_01 &  B   &                                  237 &                                            -24 &                                              754 &                                        29.2 &  2014-08-04 23   &                          3.3 &                                           -24 &                    \\
                    &        20140803\_01 &  B   &                                  218 &                                            -26 &                                              790 &                                        27.7 &  2014-08-06 20   &                          3.9 &                                           -26 &                    \\
                    &        20140923\_02 &  B   &                                  178 &                                           -102 &                                              754 &                                       -28   &  2014-09-28 09   &                          4.7 &                                          -102 &                    \\
        20160403\_01 &                    &  A   &                                  132 &                                              0 &                                              751 &                                       -23.2 &  2016-04-06 05   &                          3.4 &                                             0 &                    \\
        20161007\_01 &                    &  A   &                                  112 &                                            -56 &                                              478 &                                        -3.5 &  2016-10-12 14   &                          4.9 &                                           -56 &                    \\
        20161013\_01 &                    &  A   &                                   32 &                                            -64 &                                              345 &                                        -9.3 &  2016-10-19 15   &                          6.5 &                                           -64 &                    \\
        20161109\_01 &                    &  A   &                                   15 &                                            -90 &                                              314 &                                       -26.7 &  2016-11-17 17   &                          9.1 &                                           -90 &                    \\
        20161207\_01 &                    &  A   &                                   23 &                                            -63 &                                              381 &                                        12   &  2016-12-14 14   &                          7.1 &                                           -63 &                    \\
        20161222\_01 &                    &  A   &                                  187 &                                            -67 &                                              330 &                                        12.9 &  2016-12-29 12   &                          7.5 &                                           -67 &                    \\
\hline

\end{tabular}
}\label{tab:events_table}
\end{sidewaystable}

\begin{enumerate}
	\item Events with a clearly identifiable FD at MSL --- \textbf{40 ICMEs}
	\item ICMEs that might have interacted with others on their way to MSL, but still have a clear correspondance to a 
	FD at Mars (either completely separate or multiple steps) --- \textbf{5 ICMEs}
	\item ICMEs that probably interacted with others on their way to MSL, so that their FDs can not be 
	matched unambiguously (e.g. because there is only one merged FD or none at all) at MSL --- \textbf{53 ICMEs}
	\item Other events that don't show a clear FD at Mars (either the FD is too weak or the ICME missed MSL completely) 
	--- \textbf{19 ICMEs}
	\item Events where the analysis could not be applied due to a data gap or an SEP event at MSL/RAD coinciding with 
	the FD, or poor visibility in the STEREO-HI image leading to a high uncertainty of the predicted arrival time --- 
	\textbf{32 ICMEs}
\end{enumerate}

This categorization shows that for all 117 events where the HI and RAD data quality is sufficient (excluding 
category 5), 45 are clearly identifiable. So, based on the results for our sample, there is a 
\SI{39+-6}{\percent} chance that an ICME observed in STEREO-HI and predicted to arrive at MSL by the SSEF30 method data 
is actually observed through a FD at MSL/RAD that is clearly related to the ICME. This is consistent with 
the performances between \SIrange[range-phrase={\,and\,}]{12}{44}{\percent} for different locations in the inner 
heliosphere found by \citet[see their Table 1]{Moestl-2017-HelcatsHSO}.

In \SI{50+-8}{\percent} of the cases, multiple ICMEs interacted and possibly merged, making it impossible to 
unambiguously match the FD to a single ICME (Category 3). When only considering the 59 individual ICMEs with no others 
occurring in quick succession and similar direction (Categories 1 and 4), the chance that a clearly related FD is 
observed at Mars grows to \SI{68+-14}{\percent}. The uncertainties of the percentages given here were calculated 
assuming Poisson statistics ($\sigma = \sqrt{N}$).

Only the events in categories 1 and 2 will be used in the following studies, as these are the only ones where an 
arrival time at MSL could be defined. A list of these events with the relevant SSE fit results and FD onset times can 
be found in Table \ref{tab:events_table}.

14 of these 45 ICMEs were close to oppositions of Earth and Mars (cf. Figure 
\ref{fig:introduction_oppositions_cartoon}, left panel) and thus were also seen at Earth. In these cases, we also 
included the arrival times at Earth, which were derived in a similar fashion based on Forbush decreases in data from 
the South Pole neutron monitor. \add{We have also done a basic comparison of arrival times at Earth with visible shock 
and/or ICME structures in solar wind and magnetic field measurements from ACE, and found that the Forbush decrease 
onsets agree reasonably well (usually within $\sim$ $\pm\SI{2}{\hour}$) with this data.}

\subsection{Comparison of FD results with STEREO-HI predictions}
\label{sec:arrival_comparison}

\begin{sidewaysfigure}
	\centering
	\includegraphics{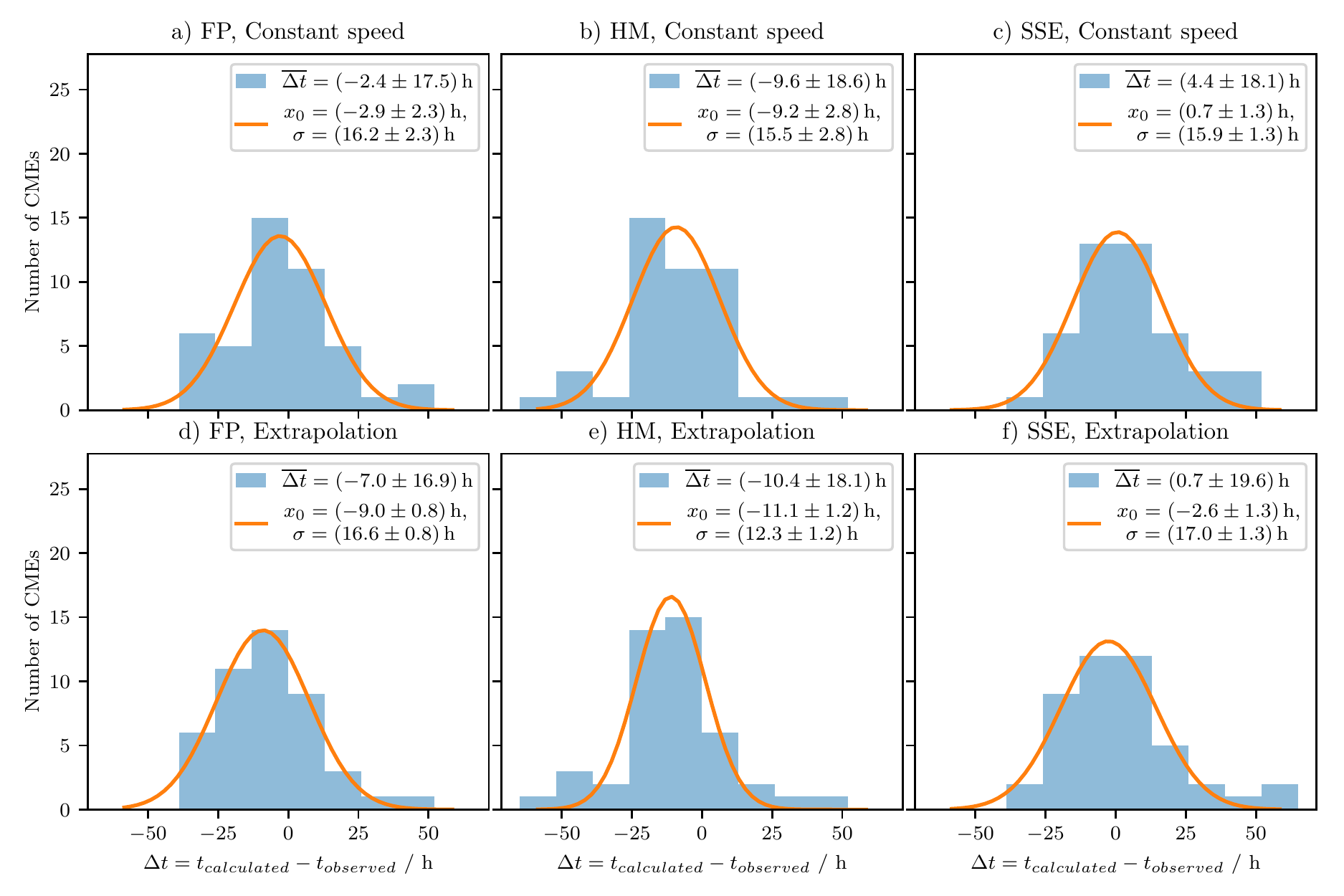}
	\caption[Comparison of arrival times predicted using STEREO-HI to the FD onset]{Comparison of arrival times 
		predicted using STEREO-HI tracking with the FP, HM and SSE methods to actual 
		FD onset times measured at MSL/RAD. \add{$\Delta t < 0$ or $\Delta t > 0$ corresponds to an earlier/later 
		predicted arrival compared to the FD onset.} The top three panels use the fitting results assuming a constant 
		speed 
		throughout the HI field of view, the lower panels extrapolate the results using a line fitted to the last 10 
		points of the $r(t)$ trajectory. All panels have the same x and y axis scaling as well as the same histogram 
		bins. Additionally, a Gaussian distribution fitted to the histograms is displayed. The legends show both the 
		results for the mean and standard deviation of $\Delta t$ calculated directly from the distribution as well as 
		the parameters $x_0$ and $\sigma$ obtained from the Gaussian fit.}
	\label{fig:arrival_time_comparison}
\end{sidewaysfigure}

For the 45 events where a clear FD at Mars was found (see Table \ref{tab:events_table}), we compared the measured FD 
onset times to the predicted arrival times using the three 
single-spacecraft fitting methods used in the HELCATS catalog. Figure \ref{fig:arrival_time_comparison} shows 
histograms of the time difference $\Delta t = t_\text{calculated} - t_\text{observed}$ for each of the three methods 
with two approaches.

First, the upper panels of Figure~\ref{fig:arrival_time_comparison} show the results from the original techniques 
\citep[summarized by][]{Moestl-2014}, which assume a constant speed and direction to calculate the CME arrival times at 
Mars. They are based on the CME speed and launch times in the HELCATS HIGeoCAT catalog and we take into account the 
appropriate correction equations for the circular CME front shapes in the HM and SSE geometries 
\citep[see][]{Moestl-2011-arrival-correction,Moestl2013-arrival-correction}. This is similar to the aforementioned 
ARRCAT for SSEF, but here we also include the FPF and HMF methods.

Secondly, in the lower three panels, results on the differences between calculated and observed arrival times are 
presented which were generated with an extrapolation method. Here, we do not assume a constant CME speed over the whole 
HI field of view, but instead only for an interval over the last 10 available HI data points, allowing an acceleration 
or deceleration before this time. The physical reason behind this assumption is that CMEs decelerate or accelerate 
closer to the Sun and eventually reach an equilibrium with the background solar wind \citep[e.g.][]{Liu-2013}. For the 
larger heliocentric distances of Mars compared to the inner planets, an assumption of constant speed over the full HI 
field of view may lead to systematic shifts in the predicted arrival times. To this end, for each event an $r(t)$ 
kinematic is created by the FP, HM and SSE conversion methods for elongation to distance, including the correction 
formulae for HMF and SSEF, and a linear fit was applied to the 10 points farthest from the Sun. The arrival time of the 
CME is then taken as the time where the linear fit intersects the current radial distance of Mars from the Sun.

Note that this second method is not fully self-consistent. In order to derive $r(t)$ we need a CME direction, and the 
direction from single-spacecraft fitting was determined with the assumption of constant speed. Also, the uncertainty in 
the elongation value of the last 10 points may be larger due to the CME appearing more faint in the HI images as it 
expands. This extrapolation method should thus be seen as a first step towards allowing a variation of the CME speed 
with HI fitting methods \citep[elaborated 
by][]{Tucker-Hood-2015-HI-Realtime,Rollett-2016-ElEvoHI,Amerstorfer-2018-ElEvoHI}, but needs to be taken with a grain 
of salt due to the abovementioned issues.

Comparing the Fixed-$\phi$, Harmonic Mean and SSE methods as shown in Figure \ref{fig:arrival_time_comparison}, all 
perform similarly well in predicting the arrival times, with FPF and HMF (panels a) and b)) giving a slightly earlier 
arrival ($\Delta t < 
0$) and SSEF (panel c)) a slightly later arrival ($\Delta t > 0$) on average. The standard deviations are also similar 
for all 
three methods, they all fall into the \SIrange{17}{20}{\hour} range. The extrapolation method in panels d), e), and f) 
also leads to almost similar results as the original fitting techniques.

Our result for SSE with the assumption of a constant speed ($\Delta t = \SI{4.4+-18.1}{\hour}$, panel c)) is also very 
similar to 
the values obtained by \citet[their Table 1]{Moestl-2017-HelcatsHSO} for arrival time predictions at other 
locations in the heliosphere: Their results lie between 
\SIrange[range-phrase={\,and\,}]{-2.1}{8.0}{\hour} for the average $\Delta t$ and between 
\SIrange[range-phrase={\,and\,}]{12.9}{23.4}{\hour} for its standard deviation. It is interesting that there seems to 
be no clear dependence of the prediction accuracy (neither for the average value nor its standard deviation) on the 
distance from the Sun, even though one would intuitively think that the 
uncertainty should increase with greater distance as errors in the model accumulate. However, at least up to Mars 
($\sim\SI{1.5}{\AU}$), this dependence is probably not 
significant compared to the inherent uncertainty coming from the strong geometric and kinematic assumptions of the 
method. This might change with new models such as the 
ElEvoHI method \citep{Rollett-2016-ElEvoHI}, which relaxes the assumption of a constant speed and also allows the ICME 
front to have an elliptical shape instead of being necessarily circular. \add{The inclusion of a drag-based model 
	\protect\citep{Vrsnak-2013} for the speed evolution is especially important --- the assumption of a constant speed 
	in 
	most previous models is very often not true due to the interaction of ICMEs with the surrounding solar wind 
	\protect\citep[as shown by many previous studies, including][]{Vrsnak-2007,Forstner-2018,Witasse-2017} as well as 
	other heliospheric structures such as stream interaction regions and other ICMEs, with which the ICME can collide 
	\protect\citep{Temmer-2012,TemmerNitta-2015,Shen-2012,Guo-2018-SeptemberEvent}.} The ElEvoHI 
implementation is currently still 
under development and has not yet been applied automatically to large datasets such as the HELCATS catalog. 
\citet{Amerstorfer-2018-ElEvoHI} have applied ElEvoHI ensemble modeling to a case study of one event and were able to 
constrain the uncertainties to less than $\pm\SI{2}{\hour}$, which is a very promising result.

To find out to what extent the differences in average values as well as standard deviations for $\Delta t$ between the 
different models are significant, we estimated the statistical error of the mean and standard deviation by fitting a 
Gaussian profile to the histogram results from Figure \ref{fig:arrival_time_comparison}, which can also be seen in the 
same figure. The standard deviations for both parameters $\sigma$ and $x_0$, calculated as the 
square root of the covariance matrix's diagonal, lie in the \SIrange{0.8}{2.3}{\hour} range --- so in fact, one could 
argue that at least with respect to the standard deviation of $\Delta t$, all single-spacecraft methods perform almost 
equally well on average, only the systematic offset ($\overline{\Delta t}$) changes depending on the method, similar to 
what was previously found by \citet{Moestl-2014} for ICMEs arriving at \SI{1}{\AU}.

\begin{figure}
	\centering
	\includegraphics{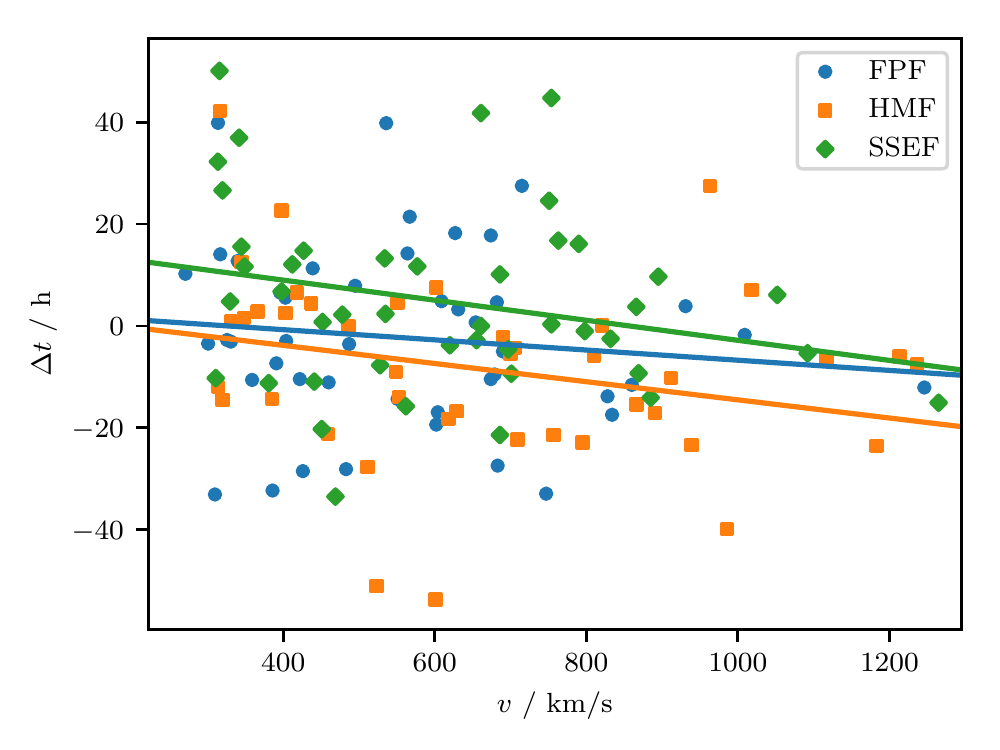}
	\caption{Plot of the dependence of the difference $\Delta t$ between the predicted and observed arrival times 
		\change{over}{on} the speed of the ICME for the three fitting methods FPF, HMF and SSEF. \add{$\Delta t < 0$ or 
		$\Delta t > 0$ corresponds to an earlier/later predicted arrival compared to the FD onset. Each of the three 
		sets of points is supplemented with a linear regression to show the general trend. The slopes of the three 
		lines are: \protect\SI{-0.010+-0.013}{\hour\per\kilo\meter\second} (FPF), 
			\protect\SI{-0.018+-0.008}{\hour\per\kilo\meter\second} (HMF), 
			\protect\SI{-0.020+-0.010}{\hour\per\kilo\meter\second} (SSEF). If we fit all three sets of points together 
			with one line to reduce the uncertainty, we obtain a slope of 
			\protect\SI{-0.019+-0.005}{\hour\per\kilo\meter\second}.}}
	\label{fig:prediction_accuracy_vs_speed}
\end{figure}

Another way to look at the predicted and observed arrival times is shown in 
Figure~\ref{fig:prediction_accuracy_vs_speed}, where $\Delta t$ is plotted against the speed of the ICME. The plot 
suggests that there is a trend to predicting an earlier arrival for fast ICMEs ($\Delta t < 0$) and a later arrival for 
slow ICMEs ($\Delta t > 0$), i.e. the speed of fast ICMEs is overestimated and the speed of slow ICMEs is 
underestimated. \add{This is supported by the linear fits shown in the figure, which have a negative slope.} 
Intuitively, this can be explained as being related to the assumption of a constant speed in the three 
fitting methods --- A fast ICME usually decelerates due to the slower surrounding solar wind, but when a constant speed 
is assumed, the predicted arrival will be earlier. This trend may be slightly less pronounced for FPF than for the 
other two methods, but in all three cases, the scatter of the data points is significant.

\add{The accuracy of the three different geometric methods can also depend on the longitudinal separation $\Delta \phi$ 
of the ICME and 
the observing STEREO spacecraft (which is also shown in Table \ref{tab:events_table}). For example, when applying the 
FP method to an event with a large longitudinal separation, the geometry can lead to a perceived acceleration of an 
ICME at larger radial distances, which is often not physical \citep{Liu-2013,Liu-2016}. However, this does not directly 
apply to the FPF, HMS and SSEF fitting methods, as in this case, $\Delta \phi$ is not pre-defined, but rather a result 
of the fitting method and therefore differs between the three methods. For this reason, we have not found a clear trend 
for the dependence of the prediction accuracy on the longitudinal separation.}

The predicted arrival time uncertainties found here with a standard deviation of \SIrange{17}{20}{\hour} for the three 
geometric models are similar to those obtained via some other approaches of modelling ICME propagation, such as the 
magnetohydrodynamic WSA-ENLIL+Cone model \citep{Odstrcil-2004} as well as the analytical Drag-Based Model 
\citep{Vrsnak-2013}. For example, \citet{Vrsnak-2014}, who applied both models to a sample of 
50 ICMEs arriving at Earth, found a standard deviation of \SI{16.9}{\hour} for ENLIL and \SIrange{18.3}{19.1}{\hour} 
for DBM. A more recent study by \citet{Dumbovic-2018-DBEM} using a different sample of 25 events and the ensemble 
modeling versions of ENLIL and DBM found the standard deviations to be slightly lower at \SI{14.4}{\hour} for ENLIL and 
\SI{16.7}{\hour} for DBM. For arrivals at Mars, these standard deviations are expected to increase slightly more, as 
the uncertainties of the models accumulate because no measurements further away from the Sun are used to constrain the 
simulation. \citet{Forstner-2018} only found a standard deviation of \SI{11}{\hour} 
between ENLIL results and measurements, but the set of events was much smaller and carefully chosen, and the simulation 
input parameters were additionally validated using the measurement at \SI{1}{\AU}, so this is not directly comparable.  
\citet{Dumbovic-2018-DBEM} also found the prediction accuracy of DBEM to be dependent on the CME speed, where fast CMEs 
are 
predicted to arrive earlier than observed --- similarly to our STEREO-HI results.

This implies that the forecasting accuracy is very similar for the different fitting methods based on STEREO-HI data 
and simulations using DBM or ENLIL. Specifically speaking, DBM and ENLIL contain a more sophisticated description of 
the ICME kinematics and its interaction with the solar wind, but their initial parameters are only based on 
observations close to the Sun. STEREO-HI, on the other hand, is a direct observation much further out into the 
heliosphere, but the simple reconstruction methods based on single-spacecraft observations need to make strong 
assumptions on the ICME geometry and kinematics.

\add{In \protect\citet{Forstner-2018}, we also applied a different method of forecasting the arrival time at Mars by 
using the in situ measurement at \SI{1}{\AU} as the inner boundary condition for the DBM model, yielding a higher 
accuracy. However, this method is only applicable when in situ measurements are available at multiple locations 
preferentially radially aligned. Such observations are unfortunately rather limited and often taken from planetary 
missions which are not optimized for studying heliospheric physics \citep{Witasse-2017,Winslow-2018,Wang-2018}. In the 
future, with more data from Parker Solar Probe and Solar Orbiter, we hope to have improved space weather forecasts 
based on modeling of CME propagations constrained by in situ measurements.}

\section{Conclusions and Outlook}\label{chp:conclusions}
In this work, we studied 149 ICMEs propagating towards Mars (between August 2012 and February 2017) as well as 
towards the MSL spacecraft during its flight to Mars (between December 2011 and July 2012). These events were observed 
remotely with the Heliospheric Imagers onboard the STEREO spacecraft as well as \textit{in situ} at their arrival 
through FDs measured with MSL/RAD. Links were established between the remote and in situ observations, 
with a \SI{39+-6}{\percent} chance that an ICME that is headed for MSL $\pm \SI{30}{\degree}$ according to the SSE 
model corresponds to a clear FD observed at RAD. This can be seen as a measure of the SSE model's performance for 
predictions of ICME arrivals at MSL's location using FDs as the method of identification of the arrival. In 
\SI{50+-8}{\percent} of cases, there was likely interaction between multiple ICMEs --- when excluding these, the chance 
of a clear and distinguishable FD grows to \SI{68+-14}{\percent}.

For the 45 remaining ICMEs where a clear FD could be observed at MSL/RAD, we could also measure the 
accuracy for predicting the arrival time of the ICME using the SSE, Fixed-$\phi$ and harmonic mean geometries. The 
average $\Delta t$ varies between 
\change{\protect\SIrange{-19.7}{+4.4}{\hour}}{\protect\SIrange[range-phrase={\,and\,}]{-10.4}{+4.4}{\hour}}, while the 
standard deviations are all very similar 
between \SIrange[range-phrase={\,and\,}]{16.9}{19.6}{\hour}. The results for the SSE method --- $\Delta t = 
\SI{4.4+-18.1}{\hour}$ --- are very comparable to the values that \citet{Moestl-2017-HelcatsHSO} found for arrivals 
at different locations in the inner heliosphere.

These standard deviations are also similar to the performance of most other current ICME modelling 
approaches, such as the WSA-ENLIL+Cone model and the Drag-Based Model. In the future, better results can probably be 
obtained with \change{even more complex}{more physical and realistic} geometric models applied to heliospheric imager 
data, such as the ElEvoHI method \citep{Rollett-2016-ElEvoHI} as well as new state-of-the-art 
MHD simulations such as the recently developed EUHFORIA model \citep{Pomoell-2018-EUHFORIA}. Certainly, more 
multi-point HI observations would also be helpful, which are currently not possible as data from STEREO B is not 
available. Future missions carrying heliospheric imagers include NASA's recently-launched Parker Solar Probe 
\citep{Fox-2016-PSP} with its WISPR instrument \citep{Vourlidas-2016-PSP-WISPR}, ESA's Solar Orbiter mission 
\citep{Mueller-2013-SolO} currently scheduled to launch in 2020 with its SoloHI \citep{Howard-2013-SoloHI}, as 
well as a possible future mission to the L5 Lagrange point \citep[proposed by 
e.g.][]{Gopalswamy-2011-L5mission,Vourlidas-2015-L5mission,Lavraud-2016,Kraft-2017-Lagrange}.

In a future study, we plan to use the catalog of ICMEs at Mars built in this work to investigate the 
properties of the Forbush decreases, such as their magnitude and steepness, in more detail. For the 14 events seen at 
both Earth and Mars, we will also directly compare these data at the two planets for better understanding their 
properties and interplanetary propagations. \add{For up to about 6 events from this catalog, the ICME and Forbush 
decrease properties can be better studied by also taking into account in situ measurements of solar wind and magnetic 
field from the MAVEN spacecraft that arrived at Mars in late 2014. However, MAVEN can only measure the upstream solar 
wind intermittently due to the spacecraft orbit (as discussed in the Introduction), resulting in frequent gaps in the 
data.} 

Nevertheless, our current results are important for the understanding of space weather, but of course also for the 
development of future deep 
space missions and human spaceflight to Mars, where accurate predictions of ICMEs, their associated shocks and 
accompanying SEP events, and their impact are essential.

\section{Sources of Data and Supplementary Material}\label{chp:data}
This section includes references to all the data used in this work.

The \textbf{HELCATS catalogs} are available from the HELCATS website, \url{https://www.helcats-fp7.eu}:\\
HIGeoCat: \url{https://doi.org/10.6084/m9.figshare.5803176.v1}\\
HIJoinCat: \url{https://www.helcats-fp7.eu/catalogues/wp2_joincat.html}\\
ARRCAT: \url{https://doi.org/10.6084/m9.figshare.4588324}

\textbf{RAD data} are archived in the NASA planetary data systems’ planetary
plasma interactions node (\url{http://ppi.pds.nasa.gov/}). Other file formats can be provided by the authors on request.

\textbf{South Pole neutron monitor} data can be retrieved from the Neutron Monitor Database at \url{http://nmdb.eu}.

A text file version of Table \ref{tab:events_table} in this work can be found on FigShare at 
\url{https://doi.org/10.6084/m9.figshare.7440245}.

\section*{Acknowledgements}

J. v. F. thanks Joana Wanger, who helped with the task of marking FD onset times in the MSL/RAD data during her 
internship in the Extraterrestrial Physics group at Kiel University.

C. M. thanks the Austrian Science Fund (FWF): [P26174-N27].

M. T. acknowledges the support by the FFG/ASAP Programme under grant no. 859729 (SWAMI).

M. D. acknowledges funding from the EU H2020 MSCA grant agreement No 745782 (project ForbMod).

J. G. is partly supported by the Key Research Program of the Chinese Academy of Sciences under grant no. XDPB11.

RAD is supported by NASA (HEOMD) under JPL subcontract 1273039 to Southwest Research Institute and in Germany by DLR 
and DLR's Space Administration grants 50QM0501, 50QM1201, and 50QM1701 to the Christian Albrechts University, Kiel.
We acknowledge the NMDB database (www.nmdb.eu), funded under the European Union's FP7 Programme (contract 213007), for 
providing data. The data from South Pole neutron monitor is provided by the University of Delaware with support from 
the U.S. National Science Foundation under grant ANT‐0838839.

\end{document}